\documentclass[letterpaper,5p,times,twocolumn]{YYYY}
\usepackage{amssymb}
\usepackage{amsmath}
\usepackage{mathtools}
\usepackage{graphicx}

\newcommand{\be}{\begin{equation}}
\newcommand{\ee}{\end{equation}}
\journal{YYYY}

\begin{document}
\begin{frontmatter}

\title{Fractional Maps and Fractional Attractors. Part I:
  $\alpha$-Families of Maps}

\author[First]{M. Edelman}

\address[First]{Dept. of Physics, Stern College at Yeshiva University, 
245 Lexington Ave, New York, NY 10016, USA; Courant Institute of
Mathematical Sciences, New York University,251 Mercer St., New York, NY
10012, USA (e-mail: edelman@cims.nyu.edu)}


\begin{abstract}                

In this paper we present a uniform way to derive families of maps from the
corresponding differential equations describing systems which
experience periodic kicks. The families depend on a single parameter - the
order of a differential equation $\alpha > 0$. We investigate 
general properties of such families and how they vary with the increase in
$\alpha$ which represents increase in the space dimension and the
memory of a system (increase in the weights of the earlier states).
To demonstrate  general properties of the $\alpha$-families we use
examples from physics (Standard $\alpha$-family of maps) and population
biology (Logistic $\alpha$-family of maps). We show that with the increase 
in $\alpha$ systems demonstrate more complex and chaotic behavior.

\end{abstract}

\begin{keyword}
Discrete map \sep Fractional dynamical system \sep Attractor 
\sep Periodic trajectory \sep Map with memory \sep Stability  
\end{keyword}

\end{frontmatter}

\section{Introduction}
\label{int}
Since the first fractional maps (FM) were derived from 
fractional differential equations (FDE)
\cite{TZ1,TarMap1,TarMap2,TarasovBook2011}, 
their investigation revealed new unusual properties
of the fractional dynamical systems \cite{ETFSM,TEdisFM,myFSM,Taieb}. 
Probably the most unusual 
feature of the investigated fractional maps is the existence of
the new type of attractors - cascade of bifurcation type trajectories (CBTT).

A cascade of bifurcations, when with the change in the value
of a parameter a system undergoes a sequence of period-doubling
bifurcations, is a well known pathway of the transition form order to
chaos (see for example \cite{AP}). 
It is associated with the Feigenbaum scaling, Feigenbaum's functional
equation, and period-doubling renormalization operator 
\cite{Feig1978,Lanford,VSKh,Cvitanovic,Briggs}. The scaling property of
dynamical systems is reflected in two Feigenbaum constants, which are
universal for large classes of systems and can be computed explicitly 
using functional group renormalization theory. 
In the case of the CBTT
the period doubling occurs without a change in a system parameter and is an
internal property of a system. The FMs are much more complicated than
the corresponding integer maps - they are maps with memory in which current
values of the map variables depend on all their previous values 
\cite{Ful,Fick1,Fick2,Giona,Gallas,Stan}. The first derived from FDEs FMs 
investigated were
two-dimensional Fractional Standard Maps (FSM) obtained from the 
fractional Universal Map derived from the
corresponding FDE. 
CBTTs were found in all of them but understanding of their
origin and the necessary and sufficient conditions of their existence is far
from being complet. The fact that memory in fractional maps decays slowly
(as a power law) does not allow to use a ``short memory'' principle 
\cite{Podlubny} and complicates the investigation of the CBTTs.   
This is why the development of a simple one-dimensional
model with CBTTs is very important.

The simplest integer one-dimensional map which has been used as a
playground for the investigation of the cascades of bifurcations in
regular dynamics is the ubiquitous Logistic Map \cite{May}. 
There were a few attempts to introduce a Logistic Map with memory. 
Bifurcation diagrams in the Logistic Maps with exponentially or geometrically 
decaying memory were considered in \cite{Giona,RASarticle} 
and memory related to 
a numerical integration of the FDE was considered in \cite{Stan}. 
The major conclusion is that memory increases stability.
The maps were not derived from FDEs and 
did not demonstrate CBTTs. It is impossible to derive a  
fractional Logistic Map (FLM) from the equation of the fractional
Universal Map introduced by Tarasov (for a review see Chapter 18 from 
\cite{TarasovBook2011}) 
or in a way similar to the way in which the Universal Map obtained by
considering a system with a periodic sequence of $\delta$-function type
pulses (kicks). In the following we derive the generalized Universal Map
(we'll call this map the Universal Map omitting for the sake of brevity 
the  word generalized and hope it won't cause any confusion)
by considering kicked systems with time delays. This allows us also to
derive the FLM in a way similar to the way in which the
regular Logistic Map can be derived. These  maps turns out to be the
simplest so far one-dimensional maps with the CBTTs.

\section{Regular (Integer) Maps}

\subsection{Universal Map (Two-Dimensional) and Standard Map}

The Universal Map 
\be 
p_{n+1} = p_n - TG(x_n), \ \  x_{n+1} = x_n + Tp_{n+1} 
\label{UM}
\ee 
can be derived in a way similar to \cite{TarMap1,TarMap2,TarasovBook2011}.  
Let's consider differential equation
\be
\ddot{x}+G(x) \sum^{\infty}_{n=-\infty} \delta \Bigl(\frac{t}{T}-(n+\varepsilon)
\Bigr)=0
\label{UMDE}
\ee
with the initial conditions:
\be
x(0)=x_0, \    \ p(0)=\dot{x}(0)=p_0,
\label{SMDEIC}
\ee
were $0 < \varepsilon < 1$. This equation is equivalent to the Volterra
integral equation of second kind:
\be
x(t)=x_0 + p_0t - \int^{t}_0 d \tau  G(x( \tau )) \sum^{\infty}_{n=-\infty}
\delta \Bigl(\frac{\tau}{T}-(n+\varepsilon)\Bigr)( t-\tau ),
\label{Volt2D}
\ee
which for $(n+\varepsilon)T<t<(n+1+\varepsilon)T$ has a solution
\be
x(t)=x_0 + p_0t -  
T\sum^{n}_{k=0} G(x( Tk+T\varepsilon)) ( t-Tk-T\varepsilon),
\label{Volt2Dx}
\ee
\be
p(t)=\dot{x}(t)= p_0 -  
T\sum^{n}_{k=0} G(x( Tk+T\varepsilon)).
\label{Volt2Dp}
\ee
With the definitions 
\be
x_{n}=x(Tn), \   \ p_{n}=p(Tn)
\label{xnpn}
\ee 
this gives for t=(n+1)T
\be
x_{n+1}=x_0 + p_0(n+1)T -   
T^2\sum^{n}_{k=0} G(x( Tk+T\varepsilon)) (n-k+1-\varepsilon),
\label{Volt2Dxn}
\ee 
\be
p_{n+1}= p_0 -  
T\sum^{n}_{k=0} G(x( Tk+T\varepsilon)).
\label{Volt2Dpn}
\ee
Taking the limit  $\varepsilon  \rightarrow 0$ and taking into account
continuity of $x(t)$ ($\dot{x}(t)=p(t)$ is finite), we obtain
\be
x_{n+1}=x_0 + p_0(n+1)T -   
T^2\sum^{n}_{k=0} G(x_k) ( n-k+1),
\label{UMapx}
\ee 
\be
p_{n+1}= p_0 -  
T\sum^{n}_{k=0} G(x_k),
\label{Umapp}
\ee
which can be written in a symmetric form as a map with full memory,
where all previous states have equal weights in the definition of the
new state (see for example \cite{Stan}):
\be
x_{n+1}= x_0 +  
T\sum^{n+1}_{k=1} p_k,
\label{UmappSimx}
\ee
\be
p_{n+1}= p_0 -  
T\sum^{n}_{k=0} G(x_k).
\label{UmappSimp}
\ee
Any system with full memory can be presented as a system with one step memory 
(which sometimes is defined as a system with no memory). 
The simplest one-dimensional map with full memory
\be
x_{n+1}= \sum^{n}_{k=0} f(x_k)
\label{FullMem}
\ee 
can be written as 
\be
x_{n+1}= x_{n}+ f(x_n).
\label{NoMem}
\ee
Similarly, equations (\ref{UmappSimx}) and (\ref{UmappSimp}) can be written 
as a simple iterative area preserving
($\partial(p_{n+1},x_{n+1})/\partial(p_{n},x_{n})=1$) process with one
step memory which is
called the Universal Map:
\be
p_{n+1}= p_{n} - TG(x_n),
\label{UMp}
\ee
\be
x_{n+1}= x_{n}+ p_{n+1}T,
\label{UMx}
\ee
which is, in essence, the relationship between the values of the physical
variables on the left sides of the consecutive kicks.

The well-known Standard Map is the particular form of the Universal Map
with $G(x) = K \sin x$:
\be
p_{n+1}= p_{n} - TK \sin x, \    \ x_{n+1}= x_{n}+ p_{n+1}T.
\label{SM}
\ee

One-dimensional Logistic Map
\be
x_{n+1}=K x_{n}(1-x_{n})
\label{LM}
\ee
may formally be written in the 2D form
\be
p_{n+1}= - G_l(x_n),  \     \ x_{n+1}= x_{n}+ p_{n+1}
\label{LM2Dn}
\ee
with
\be
G_l(x)=x-Kx(1-x).
\label{GLM}
\ee
It can't be derived from Eqs.~(\ref{UMp})
and  (\ref{UMx}) but may be written as  the particular form of the
Universal Map  with two step
memory  ($n \ge 0$)
\be
p_{n+1}= p_{n} - G(x_n,x_{n-1}),
\label{UMLp}
\ee
\be
x_{n+1}= x_{n}+ p_{n+1},
\label{UMLx}
\ee
where 
\be
G(x,y)=G_l(x)-G_l(y)
\label{Gxy}
\ee
and the initial conditions 
\be
x_0=x_{-1}-G_l(x_{-1}), \     \ p_0=-G_l(x_{-1})
\label{UMLic}
\ee
with an arbitrary $x_{-1}$.

Map  (\ref{UMLp}) -  (\ref{UMLic}) may formally be derived from the
differential equation 
\be
\ddot{x}+G(x(t),x(t-1)) \sum^{\infty}_{n=-\infty} \delta (t-n-\varepsilon)
=0
\label{UMLDE2D}
\ee
with the initial conditions (\ref{UMLic}). But this equation at present has no
physical or biological justification and won't be considered in this paper.

\subsection{Logistic Map and Universal One-Dimensional Map}

If we try to use Eq.~(\ref{UMDE}) with the first derivative instead of the
second to derive the 1D analog of the Universal Map, we will encounter the
following difficulty: in the 1D analog of the Eq. (\ref{Volt2Dx}) $x(Tk+T\varepsilon)$
will be undefined because $Tk+T\varepsilon$ is a point of discontinuity for the
function $x(t)$. To overcome this difficulty let's consider the equation
with the time delay
\be
\dot{x}+G(x(t- \Delta T)) \sum^{\infty}_{n=-\infty} \delta \Bigl(\frac{t}{T}-(n+\varepsilon)
\Bigr)=0
\label{UM1Ddif}
\ee
with the initial condition:
\be
x(0)=x_0,
\label{UM1DIC}
\ee
where $0 < \varepsilon<1$ and  $0< \Delta <1$.
Then 1D analog of Eq. (\ref{Volt2Dx}) becomes
\be
x(t)=x_0  -  
T\sum^{n}_{k=0} G(x[ T(k+\varepsilon-\Delta)]).
\label{Volt1Dx}
\ee
$x[ T(k+\varepsilon-\Delta)]= x(Tk) $ because $\dot{x}=0$ for $t \in
(T(k+\varepsilon-1), T(k+\varepsilon))$.
As a result, the corresponding 1D map with full memory becomes
\be
x_{n+1}= x_0 -  
T\sum^{n}_{k=0} G(x_k),
\label{UM1Dfmem}
\ee
which is equivalent to the 1D form of the Universal Map with one step memory
\be
x_{n+1}= x_n - T G(x_n).
\label{UM1D}
\ee

Logistic Map can be obtained from  Eq. (\ref{UM1D}) by taking
\be
G(x)=\frac{1}{T}[x-Kx(1-x)].
\label{LMG}
\ee
Differential equation (\ref{UM1Ddif}) with no time delay, no delta
functions, and $G(x)$ defined by  (\ref{LMG}) is one of the most
general models in population biology and epidemiology \cite{BC,TIS}. 
Three terms in
$G(x)$ represent the growth rate proportional to the current population,
the restrictions due to the limited resources, and the death rate. 
Logistic Map appears and plays an important role not only in the life
relevant sciences but also in economics, condensed matter physics, and
some other areas of science  \cite{TIS,AD}. In many cases time delays play
an important role (see for example Ch. 3 from  \cite{TIS} and Ch. 3 from
\cite{AD}). In the simplest case it can be related to the time of the
development of an infection in a body until a person becomes infectious,
or to the time of the development of an embryo. The delta functions model
changes that occur as periodically following discrete events. 

Finally, let us notice that the absence of a time delay is not
essential in (\ref{UMDE}) considered over any finite time interval
and both the 1D and 2D Universal Maps can be derived from the equation
\be
\frac{d^ix}{dt^i}+G(x(t- \Delta T)) \sum^{\infty}_{n=-\infty} \delta \Bigl(\frac{t}{T}-(n+\varepsilon)
\Bigr)=0,  \   \  \varepsilon > \Delta > 0, \  \ i\in\{1,2\} 
\label{UM1D2Ddif}
\ee
in the limit $\varepsilon  \rightarrow 0$: the 2D Universal Map (\ref{UMp}),
(\ref{UMx}) corresponds to $i=2$ and the 1D Universal Map (\ref{UM1D})
corresponds to $i=1$.

\section{Fractional Maps}

Let's consider (\ref{UM1D2Ddif}) in which the derivative of an integer order is
substituted by a fractional order $\alpha$ derivative.

\subsection{Riemann-Liouville Universal Map}

In the case of the Riemann-Liouville fractional derivative we have
\be
_0D^{\alpha}_tx(t) +G(x(t- \Delta T)) \sum^{\infty}_{n=-\infty} \delta \Bigl(\frac{t}{T}-(n+\varepsilon)
\Bigr)=0,  \    \ \varepsilon > \Delta > 0,  
\label{UM1D2DdifRL}
\ee
 where  $\varepsilon  \rightarrow 0$, $0 \le N-1 < \alpha \le N$,  and the initial conditions 
\be
(_0D^{\alpha-k}_tx)(0+)=c_k,  \    \ k=1,...,N.
\label{UM1D2DdifRLic}
\ee
The left-sided Riemann-Liouville fractional  derivative $_0D^{\alpha}_t x(t)$ defined for
$t>0$ \cite{Podlubny,SKM,KST} as 
$$
_0D^{\alpha}_t x(t)=D^n_t \ _0I^{n-\alpha}_t x(t)=
$$
\be
\frac{1}{\Gamma(n-\alpha)} \frac{d^n}{dt^n} \int^{t}_0 
\frac{x(\tau) d \tau}{(t-\tau)^{\alpha-n+1}}  \quad (n-1 \le \alpha < n),
\label{RL}
\ee
$D^n_t=d^n/dt^n$, and $ _0I^{\alpha}_t$ is a fractional integral.

This problem for a wide class of functions G(x) can be reduced
\cite{TarasovBook2011,KST,KBT1,KBT2} to the Volterra integral equation of second
kind ($t>0$)
$$
x(t)= \sum^{N}_{k=1}\frac{c_k}{\Gamma(\alpha-k+1)}t^{\alpha -k}  
$$
\be
-\frac{1}{\Gamma(\alpha)} \int^{t}_0 d \tau \frac{G(x( \tau - \Delta T ))}{( t-\tau )^{1-\alpha}} \sum^{\infty}_{k=-\infty}
\delta \Bigl(\frac{\tau}{T}-(k+\varepsilon)\Bigr).
\label{VoltRL}
\ee

Case  $\Delta =0$, $\varepsilon >0$, $\alpha>1$ has been considered by
Tarasov in \cite{TarMap1,TarMap2,TarasovBook2011} and its  
consideration consistent with initial conditions leads to 
the equation ($t>0$)
$$
x(t)= \sum^{N}_{k=1}\frac{c_k}{\Gamma(\alpha-k+1)}t^{\alpha -k} - 
$$
\be
-\frac{T}{\Gamma(\alpha)}  
\sum^{[t/T-\varepsilon]}_{k=0} \frac{G(x( (k+\varepsilon)T  ))}{( t-(k+\varepsilon)T )^{1-\alpha}} 
\Theta(t-(k+\varepsilon)T),
\label{VoltRLeq}
\ee
where $\Theta(t)$ is the Heaviside step function
and the map
$$
x_{n+1}=  \sum^{N}_{k=1}\frac{c_k}{\Gamma(\alpha-k+1)}[T(n+1)]^{\alpha -k} 
$$
\be
-\frac{T^\alpha}{\Gamma(\alpha)}\sum^{n}_{k=0} G(x_k) (n-k+1)^{\alpha-1}.
\label{FrRLMapx}
\ee 
With the introduction 
\be
p(t)= {_0D^{\alpha-N+1}_t}x(t) 
\label{FrMom}
\ee 
and 
\be
p^{(s)}(t)= {D^{s}_t}p(t),  \    \ s=0,1,...N-2  
\label{FrMoms}
\ee 
it also leads to
$$
p^{(s)}(t)= \sum^{N-s-1}_{k=1}\frac{c_k}{(N-s-1-k)!}t^{N -s-1-k} 
$$
\be
-\frac{T}{(N-s-2)!}  
\sum^{[t/T-\varepsilon]}_{k=0} G(x((k+\varepsilon) T)  )( t-kT )^{N-s-2},  \    \ s=0,1,...N-2. 
\label{VoltRLeqp}
\ee
and
$$
p^s_{n+1}= \sum^{N-s-1}_{k=1}\frac{c_k}{(N-s-1-k)!} [T(n+1)]^{N-s -1-k} 
$$
\be
-\frac{T^{N-s-1}}{(N-s-2)!}\sum^{n}_{k=0} G(x_k) (n-k+1)^{N-s-2}.
\label{FrRLMapp}
\ee 

If momentum is defined in a usual way
\be
p(t)= D^1_tx(t), \   \ p^s(t)= D^s_tp(t),  \    \ s=0,1,...N-2,
\label{FrMomU}
\ee 
then
$$
p^s(t)= \sum^{N}_{k=1}\frac{c_k}{\Gamma(\alpha-s-k)}t^{\alpha-1-s-k}   
$$
\be
-\frac{T}{\Gamma(\alpha-1-s)}  
\sum^{[t/T-\varepsilon]}_{k=0} \frac{G(x( (k+\varepsilon)T  ))}{( t-(k+\varepsilon)T )^{2+s-\alpha}},
\label{VoltRLeqpU}
\ee
$$
s=0,1,...N-2, \    \ t\ne(k+\varepsilon)T
$$
and
$$
p^s_{n+1}= \sum^{N}_{k=1}\frac{c_k}{\Gamma(\alpha-s-k)} 
[T(n+1)]^{\alpha-1-s-k} - 
$$
\be
\frac{T^{\alpha-s-1}}{\Gamma(\alpha-s-1)}\sum^{n}_{k=0} G(x_k) 
(n-k+1)^{\alpha-s-2},  \    \ s=0,1,...N-2.
\label{FrRLMappU}
\ee 

The advantages of defining maps using  (\ref{FrMom}) and
(\ref{FrMoms})
are: a) $c_k$ in (\ref{FrRLMapp}) are defined as the initial values 
of the momentum and its derivatives; and b) the maps can be significantly
simplified for computations. For example, for $1< \alpha \le 2$ the map
equations can be written as
\be
p_{n+1}= p_n-TG(x_n),
\label{UFMalp1n2p}
\ee 
\be
x_{n+1}=\frac{T^{\alpha-1}}{\Gamma(\alpha)}\sum^{n}_{k=0} 
p_{k+1}V^1_{\alpha}(n-k+1),
\label{UFMalp1n2x}
\ee 
where
\begin{equation} \label{V1}
V^k_{\alpha}(m)=m^{\alpha -k}-(m-1)^{\alpha -k}, 
\end{equation}
and  for $2< \alpha \le 3$ as
\be
p^1_{n+1}= p^1_n-TG(x_n),
\label{UFMalp2n3p1}
\ee 
\be
p_{n+1}= T p^1_n +p_n-T^2G(x_n),
\label{UFMalp2n3p}
\ee 
\be
x_{n+1}=\frac{p_0}{\Gamma(\alpha-1)}[T(n+1)]^{\alpha-2}+\frac{T^{\alpha-1}}{\Gamma(\alpha)}\sum^{n}_{k=0} p^1_{k+1}V^1_{\alpha}(n-k+1).
\label{UFMalp2n3x}
\ee

If  $\Delta \ne 0$, then $x(t)$ can be continued on
$t \in [-\Delta,0]$. Boundness of $x(t)$ at $t=0$ requires $c_N=0$ and
$x(0)=0$. The map equations 
(\ref{FrRLMapx}) and (\ref{FrRLMapp}) are still valid in the limit  
$\varepsilon  \rightarrow 0$ ($\varepsilon > \Delta >0$). In the case $\alpha=2$ Eqs.~(\ref{FrRLMapx}) and
(\ref{FrRLMapp}) converge to  Eqs.~(\ref{UMapx}) and (\ref{Umapp}), while for 
$\alpha=1$ Eq.~(\ref{FrRLMapx}) converges to Eq.~(\ref{UM1D}). 

We may call  Eqs. (\ref{UM1D2DdifRL}) and (\ref{UM1D2DdifRLic})
with various map generating functions (MGF) $G(x)$  
Riemann-Liouville Universal Map generating equations (RLMGE). 
RLMGE generate  families of maps ($\alpha$-RL-families) defined by  
Eqs.~(\ref{FrRLMapx}) and (\ref{FrRLMapp}) with various $\alpha$.

\subsection{Standard and Logistic $\alpha$-RL-families}

Standard and Logistic $\alpha$-RL-families of maps are two of the most 
important examples.

The Standard $\alpha$-RL-family of maps (FSMRL) is generated by the 
generating function $G(x)=K \sin x$. For $\alpha \ne 1$ we assume $K>0$.   
For $T=1$ consistent with the finite
initial coordinate $x_0=0$ ($c_N=0$ but $c_{N-1} \ne 0$ and the 
initial value of $\dot{x}$ is still singular) $x$ and $p$ coordinates of the
map for $\alpha \ge 1$ can be written as
\be
x_{n+1}=  \sum^{N-1}_{k=1}\frac{c_k}{\Gamma(\alpha-k+1)}(n+1)^{\alpha -k} - 
\frac{K}{\Gamma(\alpha)}\sum^{n}_{k=0} \sin (x_k) (n-k+1)^{\alpha-1},
\label{FrRLSMx}
\ee 
$$
p_{n+1}= \sum^{N-1}_{k=1}\frac{c_k}{(N-k-1)!} (n+1)^{N -k-1} 
$$
\be
-\frac{K}{(N-2)!}\sum^{n}_{k=0} \sin (x_k) (n-k+1)^{N-2}.
\label{FrRLSMp}
\ee 
For  $\alpha < 1$ $x_n=0$ for all values of $n$. We consider this map
on a cylinder ($x \ \  {\rm mod} \ 2\pi$).

With the same assumptions as for the Standard Map above, 
the $x$ and $p$ coordinates of the Logistic $\alpha$-RL-family of 
maps with $G(x)=x-Kx(1-x)$ for $\alpha \ge 1$ can be written as 
$$
x_{n+1}=  \sum^{N-1}_{k=1}\frac{c_k}{\Gamma(\alpha-k+1)}(n+1)^{\alpha -k} 
$$
\be
-\frac{1}{\Gamma(\alpha)}\sum^{n}_{k=0} [x_k-Kx_k(1-x_k)] (n-k+1)^{\alpha-1},
\label{FrRLLMx}
\ee 
$$
p_{n+1}= \sum^{N-1}_{k=1}\frac{c_k}{(N-k-1)!} (n+1)^{N -k-1} 
$$
\be
-\frac{K}{(N-2)!}\sum^{n}_{k=0}  [x_k-Kx_k(1-x_k)]   (n-k+1)^{N-2}.
\label{FrRLLMp}
\ee 
As for the Standard Map, for $\alpha <1$ the Logistic RL-family produces
identical $0$.

\subsubsection{Standard $\alpha$-RL-family for 
$1 \le \alpha \le 2$}

\begin{figure}
\centering
\rotatebox{0}{\includegraphics[width=8.7 cm]{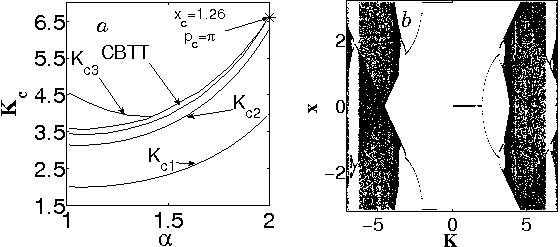}}
\caption{
Bifurcations in the Standard $\alpha$-RL-family 
of maps:
(a). Bifurcations in the fractional Standard Map for $1<\alpha<2$.  
Below $K=K_{c1}$ curve $(0,0)$ fixed point is stable. It becomes 
unstable at  $K=K_{c1}$ and gives birth to the antisymmetric $T=2$
sink which is stable for $K_{c1}<K<K_{c2}$. A pair of $T=2$ sinks with
$x_{n+1}=x_n-\pi$,  $p_{n+1}=-p_n$ is stable in the band above $K=K_{c2}$
curve. Cascade of bifurcations type trajectories (CBTTs) 
appear and exist in the narrow band which ends 
at the cusp at the top right corner of the figure. 
$(x_c,p_c)$ is the point at which the Standard
Map's ($\alpha=2$) $T=2$ elliptic points with $x_{n+1}=x_n-\pi$,  
$p_{n+1}=-p_n$ become unstable and bifurcate. In the area below $K_{c3}$
(above the CBTT band) the chaotic attractor is restricted to a band which
width is less than $2\pi$. On the upper curves and above them 
the full phase space is chaotic;
(b). Bifurcation diagram for the one-dimensional Standard Map (\ref{SM1D}). 
}
\label{figBif}
\end{figure}

The regular Standard Map ($\alpha=2$) is one of the best investigated 2D maps \cite{Chirikov,LichLib}.
Case $1< \alpha < 2$ has been investigated in a series of papers 
\cite{ETFSM,myFSM,Taieb}. Fig.~\ref{figBif}
reflects the universality of the properties of the maps belonging to the 
same family and their dependence on $\alpha$.

{\bf Low K ($K \le K_{c1}$)}. Fixed point $(0,0)$, which is a sink for   $1< \alpha < 2$
when 
\begin{equation} \label{Kc} 
K \le K_{c1}(\alpha)= \frac{2 \Gamma(\alpha)}{V_{\alpha l}},
\end{equation}
where
\begin{equation} \label{Val} 
 V_{\alpha l}  =  \sum_{k=1}^{\infty} (-1)^{k+1} V_{\alpha}^1(k),
\end{equation}
for  $ \alpha = 2$ converges to the elliptic point which is stable
for $K \le K_{c1}(2)=4$. 

The one-dimensional Standard Map ($\alpha=1$) can be written as
\be
x_{n+1}= x_n - K \sin (x_n), \ \ \ \ ({\rm mod} \ 2\pi ) ,
\label{SM1D} 
\ee
which is a particular form of the Circle Map with zero driving phase.
It has  attracting fixed points $2\pi  n$  for $0<K \le K_{c1}(1)=2$ and 
$ \pi + 2 \pi n$ when $-2 \le K < 0$ (see Fig.~\ref{figBif}b).

{\bf Antisymmetric period 2 point ($ K_{c1} \le K \le K_{c2}$}). $1< \alpha < 2$ antisymmetric period 2 sink 
\begin{equation} \label{T2AS} 
p_{n+1} = -p_n, \    \  x_{n+1} = -x_n
\end{equation} 
exists for $K>K_{c1}(\alpha)$ and is stable for 
\begin{equation} \label{T2ASstable} 
K_{c1}(\alpha) \le K \le K_{c2}(\alpha), \    \ {\rm where} \    \ K_{c2}(\alpha)= \frac{\pi \Gamma(\alpha)}{V_{\alpha l}}.
\end{equation} 
For the regular Standard Map it gives a well known result $K_{c1}(2)=4 \le K \le
K_{c2}(2)=2\pi$. For $\alpha = 1$   stability analysis produces
$K_{c1}(1)=2 \le |K| \le K_{c2}(2)=\pi$ for the stability of the period
two sink $x_{n+1}=2\pi m -x_n$.

{\bf $x_{n+1}=x_n+ \pi$ period 2 point ($ K \ge K_{c2}$ below the CBTT band)}.
For $1< \alpha < 2$ two periodic sinks
\begin{equation} \label{T2nonAS} 
p_{n+1} = -p_n, \    \  x_{n+1} = x_n+\pi
\end{equation}
appear at $K=K_{c2}$ and exist for $K \ge K_{c2}$ below the
CBTT band. For the Standard Map the corresponding elliptic points are stable 
when $2\pi \le K \le 6.59$. For $\alpha=1$  $x_{n+1} = x_n+\pi$ sinks are
stable when $\pi \le |K| \le \sqrt{\pi^2+2} \approx 3.445$.

{\bf Cascade of bifurcations band}.

\begin{figure}
\begin{center}
\rotatebox{0}{\includegraphics[width=8.7 cm]{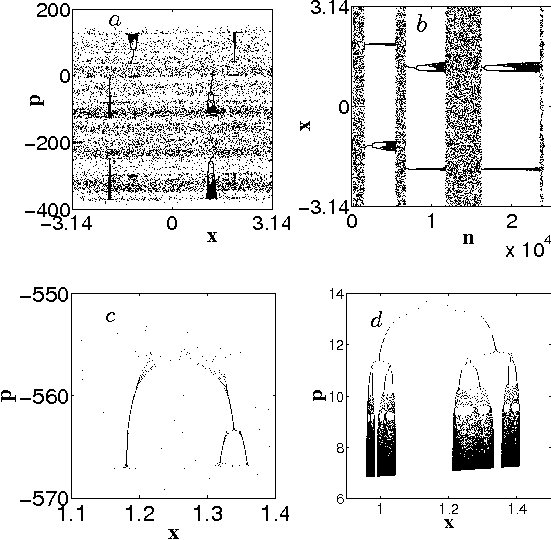}}
\caption{
Cascade of bifurcations type trajectories:
(a). $\alpha=1.65, K=4.5$; one intermittent trajectory in phase space;
(b). Time dependence of the coordinate $x$ ($x$ of $n$) for the case (a);
(c). $\alpha=1.98, K=6.46$; zoom of a small feature for a single intermittent 
trajectory in phase space;
(d). $\alpha=1.1, K=3.5$; a single trajectory enters the cascade after 
a few iterations and stays there 
during 500000 iterations. 
}
\label{Cascades}
\end{center}
\end{figure}

At $K \approx 6.59$ in the Standard Map  an elliptic-hyperbolic point
transition, 
when $T = 2$ points become unstable and stable $T = 4$ elliptic points,
appear occurs. Further increase in $K$ results in the period doubling cascade of 
bifurcations which leads to the disappearance of
the corresponding islands of stability in the chaotic 
sea at $K \approx 6.6344$. 
The cusp
in Fig.~\ref{figBif}a points to approximately this spot 
($\alpha = 2, K \approx 6.63$). Inside the
band, leading to the cusp, a new type of attractors appears: cascade of
bifurcations type trajectories (see Fig.~\ref{Cascades}). 
\begin{figure}
\centering
\rotatebox{0}{\includegraphics[width=8.7 cm]{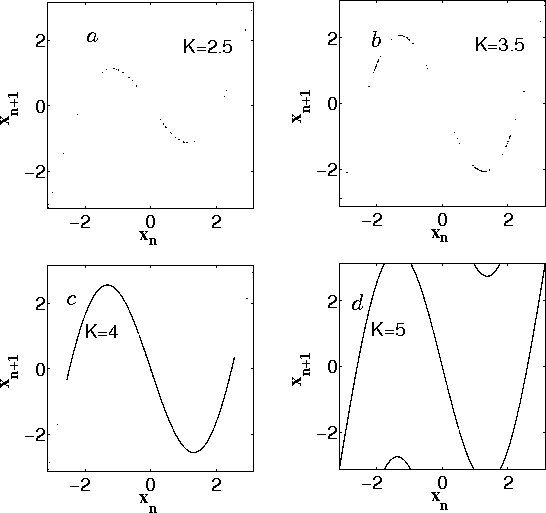}}
\caption{\label{SM1Dpp} Attractors in one-dimensional Standard Map -
$x_n$ vs. $x_{n+1}$ plots:
(a). $K=2.5$; antisymmetric $T=2$ sink;
(b). $K=3.5$; $T=4$ trajectory;
(c). $K=4$; proper attractor (width of the chaotic area is less than $2\pi$);
(d). $K=5$; improper attractor (width of the chaotic area is $2\pi$).
}
\end{figure} 
In the CBTTs period doubling
cascade of bifurcations occurs on a single trajectory without any change 
in the map
parameter. 
A typical CBTT's behavior is similar to the Hamiltonian
dynamics in the presence of sticky islands: occasionally 
a trajectory enters a CBTT and then leaves it entering
the chaotic sea (Fig.~\ref{Cascades}a,b).  
As $\alpha$ decreases the relative time
trajectories spend in CBTTs increases. 
Near the cusp (Fig.~\ref{Cascades}c) CBTTs 
are barely distinguishable and the relative
time trajectories spend in CBTTs is small.  When $\alpha$ is close to one,
a trajectory enters
a CBTT after a few iterations and stays there over the longest computational 
time
we were running our codes - 500000 iterations.
For $\alpha = 1$ sequence of bifurcations: $T=4$ at $K \approx 3.445$,  
$T=8$ at $K \approx 3.513$,
$T=16$ at $K \approx 3.526$, and so on leads to the transition to chaos at $K \approx 3.532$. 
Single antisymmetric $T=2$ ($K=2.5$),  $T=4$ ($K=3.5$), and 
two chaotic trajectories 
($K=4$ and $K=5$) are presented in Fig.~\ref{SM1Dpp}. 

{\bf $K<K_{c3}$ above the CBTT band - $x$-proper attractor}

\begin{figure}
\centering
\rotatebox{0}{\includegraphics[width=8.7 cm]{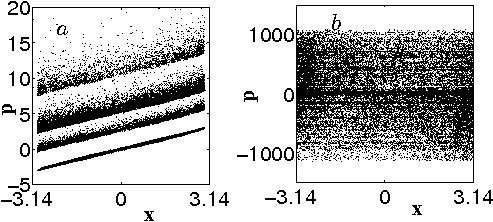}}
\caption{\label{proper} Proper and improper attractors. 3000 iterations on ten 
trajectories with the initial conditions $x_0=0$, $p_0=0.001+1.65i$, $i=0,1,...9$:
(a). A proper attractor for $K=4.2$, $\alpha=1.1$;
(b). An improper attractor for $K=4.4$, $\alpha=1.1$.
}
\end{figure}

In the one-dimensional map with $K>0$ the full phase space 
becomes involved in chaotic
motion (we'll call this case ``improper attractor'') when the maximum of 
the function $f_K(x)=x-K \sin x$ is equal to $\pi$ which
occurs at $K_{max}= 4.603339$ when $x_{max}=-1.351817$.
This is the point at which $K=K_{c3}(\alpha)$ curve on 
Fig.~\ref{figBif}a hits the $\alpha=1$ line. 
In the area between   $K=K_{c3}(\alpha)$ curve and the upper border of the 
CBTT band the fractional attractors are proper (see Fig.~\ref{proper}) 

{\bf $K=2\pi n$ accelerator mode attractor}

The regular Standard Map has a set of bands for $K$ above $2\pi n$ 
of the accelerator mode sticky islands in which momentum increases
proportionally to the number of iterations $n$ and coordinate increases as
$n^2$. The role of the accelerator mode islands (for $K$ above $2\pi$) in
the anomalous diffusion and the corresponding fractional kinetics is well
investigated (see, for example, \cite{Niyazov,ZasBook2005}). 

In the one-dimensional map the corresponding bands demonstrate  cascades of
bifurcations (see Fig.~\ref{figBif}b) for $|K|$ above $2\pi |n| $. 
The acceleration in those bands is zero and $x$ increases
proportionally to $n$.

\begin{figure}
\centering
\rotatebox{0}{\includegraphics[width=8.7 cm]{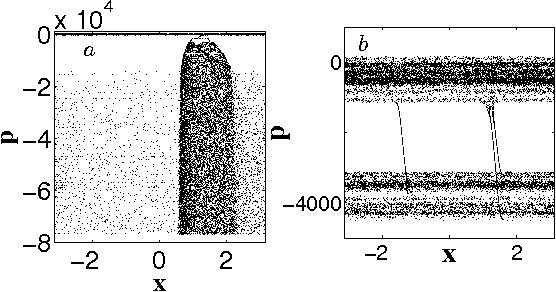}}
\caption{\label{Ballistic} Accelerator mode attractors. 25000 iterations on a
  single trajectory with the initial conditions $x_0=0$, $p_0=0.1$:
(a). CBTT-type  accelerator mode attractor for $K=5.7$, $\alpha=1.03$;
(b). Accelerator mode attractor for $K=7.6$, $\alpha=1.97$.
}
\end{figure}

The case $1<\alpha<2$ is not fully investigated. The accelerator
mode islands evolve into the accelerator mode (ballistic) attracting sticky 
trajectories when $\alpha$ is reduced from $2$ for the values of $K$ which
increase with the decrease in $\alpha$ (Fig.~\ref{Ballistic}b). When the  
value of $\alpha$ increases from 1, the corresponding ballistic attractors
evolve into the cascade of bifurcation type ballistic trajectories 
(see Fig.~\ref{Ballistic}a) for the values of  $K$ which
decrease with the increase in $\alpha$. This could mean that corresponding 
features in the one- and two-dimensional maps (at least for $K=2\pi$) 
are not connected by the continued change in $\alpha$. 

\subsubsection{Standard $\alpha$-RL-family for 
$2 \le \alpha \le 3$}
\label{alp3}

{\bf $\alpha = 3$ Universal and Standard Maps}

Following  Eqs.~(\ref{FrRLMapx}),(\ref{VoltRLeqp}), and (\ref{FrRLMapp}),
and assuming $T=1$ the three-dimensional universal map can 
be written as ($y=p$, $z=\dot{p}$) 
\be
x_{n+1}=  \frac{z_0}{2}(n+1)^2 + y_0(n+1)+x_0
-\frac{1}{2}\sum^{n}_{k=0} G(x_k) (n-k+1)^{2},
\label{3DSMx}
\ee 
\be
y_{n+1}=  z_0(n+1) + y_0 -
\sum^{n}_{k=0} G(x_k) (n-k+1),
\label{3DSMp}
\ee 
\be
z_{n+1}=   z_0 -
\sum^{n}_{k=0} G(x_k),
\label{3DSMpp}
\ee 
which can be reduced to 
$$
x_{n+1}=  x_n-\frac{1}{2}G(x_n)+y_n+\frac{1}{2}z_n, 
$$
$$
y_{n+1}=-G(x_n)+y_n+z_n,  
$$
\be
z_{n+1}=-G(x_n)+z_n,
\label{3DUM}
\ee 
or
$$
x_{n+1}=  x_n+y_{n+1}-\frac{1}{2}z_{n+1}, 
$$
$$
y_{n+1}=y_n+z_{n+1},  
$$
\be
z_{n+1}=-G(x_n)+z_n,
\label{3DUMn}
\ee 
which is a volume preserving map.

Let's assume that this map has a fixed point $(x_0,y_0,z_0)$. 
Then $z_0=y_0=G(x_0)=0$ and
stability of this point can be analyzed considering the eigenvalues of the
matrix (corresponding to the tangent map)
\[ \left( \begin{array}{ccc}
1-0.5\dot{G}(x_0) & 1 & 0.5 \\
-\dot{G}(x_0) & 1 & 1 \\
-\dot{G}(x_0) & 0 & 1 \end{array} \right).\]
The only case in which the fixed point is stable is  $\dot{G}(x_0)=0$,
when $\lambda_1=\lambda_2=\lambda_3=1$.

For the three-dimensional Standard Map with $G(x)=K \sin(x)$ 
$$
x_{n+1}=  x_n+y_{n+1}-\frac{1}{2}z_{n+1}, \   \  ({\rm mod} \ 2\pi ),
$$
$$
y_{n+1}=y_n+z_{n+1},  \   \  ({\rm mod} \ 2\pi ),
$$
\be
z_{n+1}=-K\sin(x_n)+z_n, \   \ ({\rm mod} \ 4\pi )  
\label{3DSMn}
\ee 
it means that 
fixed points $(2\pi n, 2\pi m, 4 \pi k)$ and $(2\pi n+ \pi , 2\pi m, 4 \pi k)$ , 
$n \in  \mathbb{Z}$, $m \in  \mathbb{Z}$, $k \in  \mathbb{Z}$, 
are unstable for all $K \ne 0$. Ballistic points $K\sin(x)=-4\pi n$,
$y=2\pi m$, $z=4\pi k$, which appear for $|K| \ge 4\pi$, are also unstable.   


Stability of $T=2$ points is defined by the eigenvalues of the matrix 
\[ \left( \begin{array}{ccc}
1-0.5K \cos x_1 & 1 & 0.5 \\
-K \cos x_1 & 1 & 1 \\
-K \cos x_1 & 0 & 1 \end{array} \right) \times
\left( \begin{array}{ccc}
1-0.5 K \cos x_2 & 1 & 0.5 \\
-K \cos x_2 & 1 & 1 \\
-K \cos x_2 & 0 & 1 \end{array} \right).\]
For the stable (on the torus) period two point   
$$
z_1,\  \ y_1=\frac{z_1}{2}-\pi(2n+1),\  \ K \sin x_1=2z_1,  
$$
\be
 z_2=-z_1, \  \ y_2=-\frac{z_1}{2}-\pi(2n+1),\  \ x_2=x_1-\pi(2n-1),  
\label{3DT2StablePoint}
\ee 
where $n \in  \mathbb{Z}$, the  eigenvalues are
\be
 \Bigl\{1, \  \  \frac{1}{8}(8-K^2 \cos^2 x_1 \pm K \cos x_1 \sqrt{K^2\cos^2
  x_1-16})  \Bigr\}
\label{3DT2StablePointEigenV}
\ee 
\begin{figure}
\centering
\rotatebox{0}{\includegraphics[width=8.7 cm]{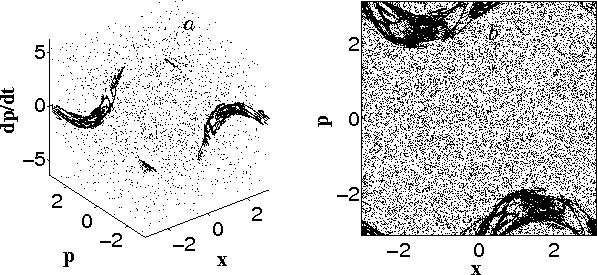}}
\caption{\label{3D} Phase space of the 3D Standard Map (\ref{3DSMn}) for $K=3$:
(a). Phase space in three dimensions; (b). Projection of the phase space
on the $x$-$y$ plane.   
}
\end{figure} 
Stable $T=2$ points exist along a line defined by
Eqs.~(\ref{3DT2StablePoint}) for all values of $z$ satisfying the condition
\be
K^2-16<4z^2<K^2.
\label{3DT2zCond}
\ee
An example of phase space for $K=3$ in three dimensions and its projection
on the $x$-$y$ plane is given in Fig.~\ref{3D}. In this case  $T=2$ points
are stable for $-1.5<z<1.5$
and the space around the line of stability presents a series of islands
(invariant curves), islands around islands, and separatrix layers. 
As $K$ goes to zero, the
volume of the regular motion shrinks. For small values of $K$ the line of the 
stable $T=2$ points exists for $-K/2<z<K/2$. 
A different form of the 3D volume preserving Standard Map was introduced
and investigated in details in \cite{DM}.

{\bf $2< \alpha \le 3$}

Assuming $G(x)=K \sin (x)$ and
$T=1$ in Eqs.~(\ref{UFMalp2n3p1})-(\ref{UFMalp2n3x}), the FSMRL for 
$2< \alpha \le 3$ can be written as

\be
p^1_{n+1}= p^1_n-K\sin(x_n),
\label{SMRLalp2n3p1}
\ee 
\be
p_{n+1}= p^1_n +p_n-K\sin(x_n), \   \  ({\rm mod} \ 2\pi ),
\label{SMRLMalp2n3p}
\ee 
\be
x_{n+1}=\frac{p_0}{\Gamma(\alpha-1)}(n+1)^{\alpha-2}+\frac{1}{\Gamma(\alpha)}\sum^{n}_{k=0} p^1_{k+1}V^1_{\alpha}(n-k+1),
\label{SMRLalp2n3x}
\ee 
$$
 \   \  ({\rm mod} \ 2\pi ).
$$
\begin{figure}
\centering
\rotatebox{0}{\includegraphics[width=8.7 cm]{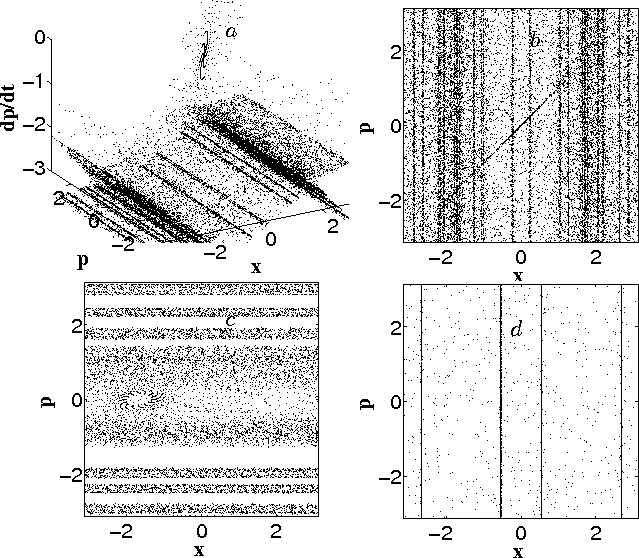}}
\caption{\label{RL2_3D} FSM for $2< \alpha <3$:
(a). 3D phase space for $K=1$, $\alpha=2.01$ obtained on a single trajectory
with $x_0=p_0=0$ and $p^1_0=0.01$; (b). Projection of the phase space in (a)
on the $x$-$y$ plane; (c). Projection of the phase space for  $K=0.2$, 
$\alpha=2.01$, $x_0=p_0=0$  
on the $x$-$y$ plane obtained using 20 trajectories with different 
initial values  of $p^1_0$; (d). The same as in (c), but for  $K=4$ and 
$\alpha=2.9$.   
}
\end{figure}
In our simulations we did not find a stable fixed point even for small
values of $K$ (see Figs.~\ref{RL2_3D} (c). 
Simulations show that for this map there are attractors in the form of 
the attracting multi-period lines with constant $x$ (see
Fig.~\ref{RL2_3D} (a), (b), and (d)). For most of the values of the map 
parameters the phase space is highly chaotic.  
   
This case and the transition from the 2D Standard Map to the 3D Standard Map, 
as well as 3D Standard Map, is not yet fully investigated.

\subsubsection{Logistic $\alpha$-RL-family for 
$1 \le \alpha \le 3$}

Case $\alpha=1$ is probably the best investigated map - the regular Logistic
Map. For $1 < \alpha \le 2$ it can be written as

\be
p_{n+1}= p_n+Kx_n(1-x_n)-x_n,
\label{LFMalp1n2p}
\ee 
\be
x_{n+1}=\frac{1}{\Gamma(\alpha)}\sum^{n}_{k=0} 
p_{k+1}V^1_{\alpha}(n-k+1),
\label{LFMalp1n2x}
\ee 
which for $\alpha=2$ can be reduced to
\be
p_{n+1}= p_n+Kx_n(1-x_n)-x_n,
\label{LFMalp2p}
\ee 
\be
x_{n+1}= x_n + p_{n+1}.
\label{LFMalp2x}
\ee 
This map is similar to the original area preserving quadratic map 
considered by H\'enon
\cite{Henon69} (for a recent review see book \cite{ZeraS2010}). As for 
the Standard $\alpha$-RL-family, for the Logistic $\alpha$-RL-family 
of maps we are interested
in the stable fixed and periodic points in order to follow their evolution
as a function of $\alpha$. For $-3<K<1$ $(0,0)$ fixed point of the 
2D Logistic Map is stable. At $K=1$ it becomes unstable but fixed point 
$((K-1)/K,0)$ becomes stable. This point is stable for  $1<K<5$. 
At $K=5$ a bifurcation occurs and this stable fixed point changes to a couple
of $T=2$ stable points 
\be
x = \frac{K+3 \pm \sqrt{(K+3)(K-5)}}{2K},  \  \ p=\pm \frac{\sqrt{(K+3)(K-5)}}{2K}.
\label{LFMalp2T2}
\ee      

For $2< \alpha \le 3$ the map can written as
\be
p^1_{n+1}= p^1_n+Kx_n(1-x_n)-x_n,
\label{LFMalp2n3p1}
\ee 
\be
p_{n+1}=  p^1_n +p_n+Kx_n(1-x_n)-x_n,
\label{LFMalp2n3p}
\ee 
\be
x_{n+1}=\frac{p_0}{\Gamma(\alpha-1)}(n+1)^{\alpha-2}+\frac{1}{\Gamma(\alpha)}\sum^{n}_{k=0} p^1_{k+1}V^1_{\alpha}(n-k+1),
\label{LFMalp2n3x}
\ee 
which for $\alpha=3$ can be reduced to
$$
x_{n+1}=  x_n+y_{n+1}-\frac{1}{2}z_{n+1}, 
$$
$$
y_{n+1}=y_n+z_{n+1},  
$$
\be
z_{n+1}=Kx_n(1-x_n)-x_n+z_n.
\label{3DLMn}
\ee 
3D quadratic volume preserving  maps were investigated in 
\cite{Moser1994,LoMeiss1998}. Everything stated in Subsection \ref{alp3}
for the 3D Universal Map is still valid for the 3D Logistic Map.
More on the  Logistic $\alpha$-RL-family of
maps will be presented in the part II of the article. 

\subsection{Caputo Universal Map}

The Logistic and Standard $\alpha$-RL-families give identical zeros in the 
case  $\alpha < 1$. This is not the case for the Logistic and Standard 
$\alpha$-Caputo-families (FLMC and FSMC), 
for which the condition $x_0=0$ can be removed.
In order to get some insight into the origin of one of the most
interesting features of fractional maps - CBTTs, we complete this part of
the paper with the analysis of the simplest case of CBTTs 
in the Logistic and Standard $\alpha$-Caputo-families of the maps when    
$\alpha < 1$. But let's first introduce the Caputo Universal Map.

Similar to (\ref{UM1D2DdifRL})
in the case of the Caputo fractional derivative we have
\be
_0^CD^{\alpha}_tx(t) +G(x(t- \Delta T)) \sum^{\infty}_{n=-\infty} \delta \Bigl(\frac{t}{T}-(n+\varepsilon)
\Bigr)=0,  \    \ \varepsilon > \Delta > 0,  
\label{UM1D2DdifC}
\ee
 where  $\varepsilon  \rightarrow 0$, $0 \le N-1 < \alpha \le N$,  and the initial conditions 
\be
(D^{k}_tx)(0+)=b_k,  \    \ k=0,...,N-1.
\label{UM1D2DdifCic}
\ee
The left-sided Caputo fractional  derivative $_0^CD^{\alpha}_t x(t)$ defined for
$t>0$ \cite{Podlubny,SKM,KST} as 
$$
_0^CD^{\alpha}_t x(t)=_0I^{n-\alpha}_t \ D^n_t x(t) =
$$
\begin{equation}
\frac{1}{\Gamma(n-\alpha)}  \int^{t}_0 
\frac{ D^n_{\tau}x(\tau) d \tau}{(t-\tau)^{\alpha-n+1}}  \quad (n-1 <\alpha \le n).
\label{Cap}
\end{equation}

This problem is equivalent to the following Volterra integral equation
($t>0$) \cite{TarasovBook2011,KST}
$$
x(t)= \sum^{N-1}_{k=0}\frac{b_k}{k!}t^{k} 
$$
\be
-\frac{1}{\Gamma(\alpha)} \int^{t}_0 d \tau \frac{G(x( \tau - \Delta T ))}{( t-\tau )^{1-\alpha}} \sum^{\infty}_{k=-\infty}
\delta \Bigl(\frac{\tau}{T}-(k+\varepsilon)\Bigr).
\label{VoltC}
\ee
\begin{figure}
\centering
\rotatebox{0}{\includegraphics[width=8.7 cm]{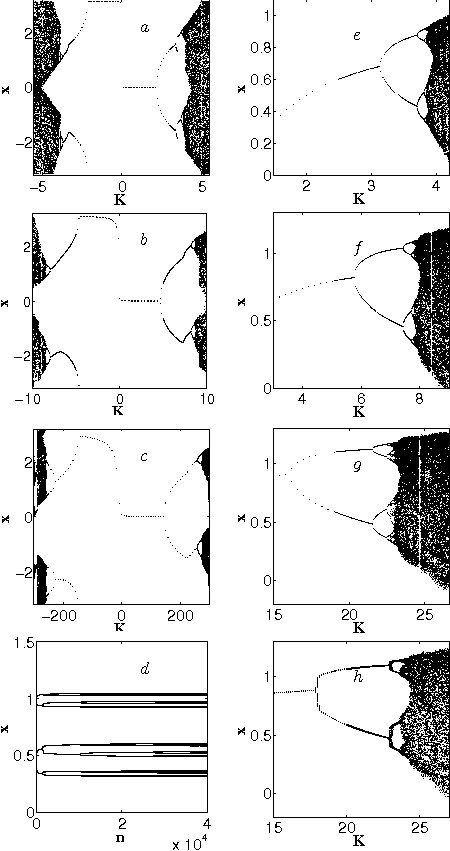}}
\caption{\label{LMandSMCap} Cascades of bifurcations and CBTTs in Standard
and Logistic $\alpha$-Caputo-families (FSMC and FLMC) for $0<\alpha<1$:
(a). Cascade of bifurcations for the FSMC (\ref{FrCMapSM})  with $\alpha=0.8$ 
obtained after performing 10000 iterations on a single trajectory
with $x_0=0.1$ for various values of $K$; 
(b). The same as in (a) but for $\alpha=0.3$; 
(c). The same as in (a) but for $\alpha=0.01$; 
(d). A single CBTT for the FLMC with  $\alpha=0.4$, $K=6.075$, and the 
initial condition $x_0=0.1$; 
(e). The same as in (a) but for the FLMC (\ref{FrCMapLM});
(f). The same as in (b) but for the FLMC; 
(g). The same as in (e) but for $\alpha=0.1$;
(h). The same as in (g) but after 100 iterations. 
}
\end{figure}

Then we also have for $t>0$
$$
x(t)=  \sum^{N-1}_{k=0}\frac{b_k}{k!}t^{k} 
$$
\be
-\frac{T}{\Gamma(\alpha)}  
\sum^{[t/T-\varepsilon]}_{k=0} \frac{G(x( (k+\varepsilon)T  ))}{( t-(k+\varepsilon)T )^{1-\alpha}} 
\Theta(t-(k+\varepsilon)T)
\label{VoltCeq}
\ee
and the map
$$
x_{n+1}= \sum^{N-1}_{k=0}\frac{b_k}{k!}[T(n+1)]^{k} 
$$
\be
-\frac{T^\alpha}{\Gamma(\alpha)}\sum^{n}_{k=0} G(x_k) (n-k+1)^{\alpha-1},
\label{FrCMapx}
\ee 
which for $0<\alpha<1$, assuming $T=1$, gives
\be
x_{n+1}=  x_0- 
\frac{1}{\Gamma(\alpha)}\sum^{n}_{k=0} G(x_k) (n-k+1)^{\alpha-1}.
\label{FrCMapxlt1}
\ee

\subsubsection{Standard and Logistic $\alpha$-Caputo-families 
for $0<\alpha<1$ }

For  $0<\alpha<1$ the Standard $\alpha$-Caputo-family can be written as 
\be
x_{n}=  x_0- 
\frac{K}{\Gamma(\alpha)}\sum^{n-1}_{k=0} \frac{\sin{x_k}}{(n-k)^{1-\alpha}},
 \   \  ({\rm mod} \ 2\pi )
\label{FrCMapSM}
\ee 
and the Logistic $\alpha$-Caputo-family as
\be
x_{n}=  x_0+ 
\frac{1}{\Gamma(\alpha)}\sum^{n-1}_{k=0} \frac{Kx_k(1-x_k)-x_k}{(n-k)^{1-\alpha}}.
\label{FrCMapLM}
\ee 
As one may see, from (\ref{FrCMapSM}), (\ref{FrCMapLM}), and 
Figs.~\ref{LMandSMCap} (a)-(c) and (e)-(g)  
a decrease in $\alpha$ and corresponding decrease in the weights of the
earlier states (decrease in the memory effects) leads to the shift of the
bifurcation curves to the higher values of the maps' parameter $K$, which means
increased stability. When $\alpha \rightarrow 1$ the bifurcation curves
converge to the curve in Fig.~\ref{figBif}b for the FSM and to the well known
bifurcation curve for the  Logistic Map in the case of the FLM.  

For the values of $K$ at which periodic $T>2$ stable points exist the
individual trajectories are the CBTTs (see Fig.~\ref{LMandSMCap}d), where an
increase in the number of the map iterations leads to the change in the
map's stability properties.  These are the simplest so far examples of the
maps with the CBTTs. 
As a result, the FSMC and the FLMC can be used as the simplest models 
for studying the CBTT phenomenon. 

The presence of CBTTs leads to the dependence 
of the the bifurcation curve on the number of
iterations. From   Figs.~\ref{LMandSMCap} (g) and (h) one may see that with
the increase in the number of iterations in the the FLMC the bifurcation
curve becomes more pronounced and shifts to the left 
(to the lower values of the map parameter $K$).


\section{Conclusion}
     
Differential equations with the integer order derivatives and the terms 
containing
periodic delta-function shaped kicks are frequently used in science to
model the behavior of real systems. They can be easily integrated
over an integer number of periods in order to obtain maps, which then can
be numerically investigated in order to derive the general properties of the
original systems. Well known examples of such maps are the Standard
Map in physics and the Logistic Map in population biology. In this paper we
proposed  continuation of the integer maps into the area of the fractional maps
corresponding to the arbitrary positive values 
of the orders of the fractional derivatives in the
original differential equations. We introduced the notions of the
Riemann-Lioville and Caputo $\alpha$-families of maps, which depend on a
single parameter - order of the fractional derivative $\alpha >0$. 
Increase in $\alpha$ corresponds to the increase in the dimension of the
space (which can be important in physical systems) and to the increase in
the memory effects (the weights of the earlier states increase with the
increase in $\alpha$), which is important in population biology 
(information about past states of the system can be recorded in the DNA
of individuals, the society as a whole can regulate evolution of
the  population by introducing laws based on the past experience).

Our first results show that increase in $\alpha$ leads to more a more
complex behavior of the system. As $\alpha \rightarrow 0$ any memory
of the past disappear and the system stabilizes at zero for the
Riemann-Lioville families or at constant value for the Caputo families. 
For $0<\alpha<1$ the behavior of the  systems is qualitatively similar to
the behavior of the system with $\alpha=1$ with stability reducing as
$\alpha \rightarrow 1$. Within the bifurcation ($T>2$) band of the
parameter values fractional systems evolve as  CBTTs.

For $1<\alpha<2$ the systems become two-dimensional and their phase space
contains different kinds of arttractors like sinks, chaotic attractors, CBTTs,
and intermittent CBTTs.      

Case $2<\alpha<3$ requires more thorough investigation. Our preliminary
results show the existence of very complicated three-dimensional attractors and
instability of the fixed points. In general, as $\alpha$ increases, the
systems become more and more chaotic.  

In the second part of this paper we'll present more results on the
fractional Logistic Map with $\alpha>1$, the fractional Standard Map with 
$\alpha>2$, and the CBTTs.

\section*{Acknowledgments}
The author expresses his gratitude to V.~E. Tarasov for useful remarks, to E. Hameiri and H. Weitzner 
for the opportunity to complete this work at the Courant Institute.

\end{document}